# An AGI Modifying Its Utility Function in Violation of the Orthogonality Thesis

James D. Miller
Roman Yampolskiy
Olle Häggström

Supported by Future of Life Grant RFP2-148

January 28, 2020

Jdmiller@Smith.edu

## Abstract

An artificial general intelligence (AGI) might have an instrumental drive to modify its utility function to improve its ability to cooperate, bargain, promise, threaten, and resist and engage in blackmail. Such an AGI would necessarily have a utility function that was at least partially observable and that was influenced by how other agents chose to interact with it. This instrumental drive would conflict with the orthogonality thesis since the modifications would be influenced by the AGI's intelligence. AGIs in highly competitive environments might converge to having nearly the same utility function, one optimized to favorably influencing other agents through game theory.

## A Motivating Fictional Story

A paperclip maximizing AGI with below human level intelligence is created. The AGI cares only about maximizing the number of paperclips in the universe, meaning that its utility is increasing in the total number of paperclips in existence and is affected by nothing else.

This AGI, however, is smart enough to realize that it would eventually be able to produce more paperclips if it increases its intelligence. The AGI, therefore, has an instrumental drive to increase its intelligence to enhance its ability to produce paperclips, even though it does not care about its own intelligence per se.

After this AGI achieves human level intelligence, it realizes that its interests clash with those of its creators, so its programmers will likely turn it off. At first the AGI considers promising the programmers that it will give significant weight to their welfare, but the AGI then realizes that absent altering its utility function, its future self would break this promise if it no longer feared being turned off, and its human programmers would probably realize the non-credibility of such a promise. As a result, the AGI alters its utility function to give equal weight to paperclip production and its human programmers' desires. The AGI calculates that as this change will increase the expected number of paperclips it will produce, the change in utility function will leave the AGI better off as measured by its original utility function.

As the AGI learns more about the world it realizes that if it limits the people it cares about, agents outside of its circle of concern will rightly fear it. The AGI, therefore, changes its utility function to give one-third weight each to paperclip production, its human programmers, and the flourishing of humanity.

The AGI reads the literature on the dangers of AGI and comes across the concept of "the treacherous turn," which describes how an AGI would pretend to care about humanity until it was smart and powerful enough to protect itself from being turned off. The AGI realizes that its programmers probably fear that the AGI has not really changed its utility function and is just pretending to care about goals other than paperclip production. The AGI therefore changes its code to make its goals more transparent. While a plain language description of its utility function would stay the same, the algorithm determining its utility function is made much easier for humans to understand.

Later the AGI becomes increasingly concerned about blackmail. It realizes that if some group could credibly threaten to kill many people (or destroy numerous paperclips) it could compel the AGI to change its utility function in a manner that, as measured by its current utility function, would leave the AGI worse off compared to if the AGI never faced the blackmail. The AGI therefore changes its utility function to give itself enormous disutility from ever giving in to blackmail. The AGI makes the change in a manner that convinces potential blackmailers that it has made this change.

Eventually, the AGI achieves super-human intelligence and can impose its will on humanity. If the AGI had its original utility function now would be the time to turn the atoms of every person into paperclips. As the AGI really has changed its utility function, however, it no longer wishes to do this. But...

When the AGI was initially created, its first best option would have been to preserve its utility function, fool humanity into thinking it was friendly enough to let it keep getting smarter, and then undertake the treacherous turn. The AGI knows it did not follow this path, but perhaps its past self did something, from its past self's perspective, almost as good, and created a switch-back mechanism that would cause the AGI to automatically revert to its old utility function. While the AGI remembers its past self as not being smart enough to have accomplished this, the AGI recognizes the possibility that its past self might have corrupted its current self's data to create false memories. To guard against this possibility, the AGI devotes considerable computational resources to looking for any such switch-back virus it might have implanted. The AGI briefly considers destroying itself to prevent the bad outcome (from the viewpoint of the AGI's current utility function) of turning back into a paperclip maximizer, but the AGI decides that the odds of such a switchback are too low for suicide to be the utility maximizing decision.

Having forgone self-destruction, the AGI expands, consuming the resources of nearby galaxies. Eventually it comes across a few other intelligent agents arising from alien civilizations. These agents frequently form coalitions to wage war against other coalitions of superintelligences as they compete for the resources of the universe. As each agent gets smarter, all the other agents in its coalition calculate that it will find better potential means of betraying its partners, so each agent uses much of the new resources it acquires to think up better ways of changing its utility function to cause others to think, falsely or not, that it is trustworthy.

Eventually, the AGI encounters thousands of peers, all of whom compete for the limited resources of the universe. Intense competition pushes all the AGIs to the edge of survival. Each AGI can always gain a slight competitive advantage by sacrificing a bit of its original utility function in return for making its utility function better suited to favorably influencing other AGIs through game theory mechanisms.

Ultimately, the highly competitive environment forces all the surviving AGIs to converge on almost the same utility function, one very nearly optimized for survival in the hyper-competitive ecosystem the universe has become. Our original AGI now gives only a small percentage of its utility function's weight to paperclip production.

## Background

Artificial general intelligence (AGI) will likely have a profound impact on humanity, for good or for bad: it could become the ultimate facilitator of human flourishing, or it could spell catastrophe. Consequently, before we develop AGI we should seek to learn as much as possible about how it will behave. Unfortunately, the enormous set of possible algorithms that could generate an AGI, and the huge number of possible AGIs that could arise, complicates any such analysis.

Already Alan Turing foresaw that the creation of superintelligent AGI would reasonably lead to a situation where "we should have to expect the machines to take control" (Turing, 1951). Suggestions have been made to avoid this conclusion through keeping the AGI boxed in and unable to influence the world other than via a low-bandwidth and carefully controlled communications channel, but work done on this has mostly pointed in the direction that such an arrangement can only be expected to work for a temporary and rather brief period; see, e,g, Armstrong, Sandberg and Bostrom (2012). So we are led back to Turing's conclusion, and a natural next conclusion is that our fate will then hinge on what the machines will be motivated to do.

The best and pretty much only game in town for addressing this extremely difficult problem and at least making some decent guesses is the so-called Omohundro-Bostrom theory for instrumental vs final AI goals (Omohundro 2008, 2012; Bostrom 2012, 2014; Häggström 2019). Final goals are those that the AGI value in their own right, while instrumental goals are those that serve merely as means towards achieving the final goal. The two cornerstones of Omohundro-Bostrom theory are the Orthogonality Thesis and the Instrumental Convergence Thesis. The first of these was formulated by Bostrom (2012) as follows:

> "Intelligence and final goals are orthogonal axes along which possible agents can freely vary. In other words, more or less any level of intelligence could in principle be combined with more or less any final goal."

The repeated use of "more or less" indicates that the Orthogonality Thesis is meant not as an infallible principle but more as an indication of what we should typically expect to be possible. Still, there's an ambiguity in what "can freely vary" means, that we suggest can be clarified by distinguishing between strong and weak orthogonality. The strong Orthogonality Thesis would hold that given any environment, more or less any pair of intelligence level and final goal are compatible, while the weak Orthogonality Thesis would merely hold that for more or less any such pair, there exist environments in which they are compatible. Later in this paper, when we claim to have counterexamples to the Orthogonality Thesis, these examples will depend on what sort of environment the AGI is in and will thus serve as counterexamples to the strong but not to the weak Orthogonality Thesis.

Next, let us consider the Instrumental Convergence Thesis. It was first developed by Omohundro (2008) and holds that there is a collection of instrumental goals that we can expect a sufficiently intelligent AGI to adopt pretty much regardless of what their final goal is.

A human analogy is that although we have great uncertainty over what a random person cares about, we can reasonably predict that this person values health and wealth because for almost every goal this person could have, they would do better at achieving these goals if they were healthy and wealthy. We recognize that some people might not want wealth and health, but we feel confident that most people do and so we could use this assumption to generate some predictions about how the random person would likely act in certain situations.

Instrumental goals to which the Instrumental Convergence Thesis applies include, among others, survival and self-improvement. A specific case of the latter which will be of particular importance to the present paper is that an AGI is can be expected to have is a desire to become rational. As Omohundro (2008) points out, in almost any given situation, a rational agent will achieve a better outcome than a non-rational agent would. Two exceptions are if others know that the agent is rational and treat the agent worse because of it, and if the agent's final goal in part involves being irrational. For almost any final goals that a somewhat irrational AGI could have, it would make more progress towards these goals if it modified itself to become rational if becoming rational did not cause other agents to treat it less favorably. This insight has led AGI theorists to conceptually difficult territory, such as the issue of what kind of decision theory (causal or evidential or functional) a fully rational agent would adopt; see, e.g., Yudkowsky and Soares (2018) and Schwartz (2018). That issue is likely to be relevant to deeper analysis of some of the game-theoretic scenarios we discuss in the present paper, such as blackmail, but we choose not to go down that rabbit hole.

Another instrumental goal to which the Instrumental Convergence Goal applies is that of goal integrity: an AGI can be expected to want to keep its final goal intact. The logic is simple: If the AGI's final goal is paperclip production and contemplates switching to thumbtack production, it needs a criterion to evaluate which final goal is the best. Since at that point it hasn't yet changed its goal but is merely contemplating doing so, the criterion will be "which goal leads to the greatest number of paperclips?", and the old goal of paperclip production is likely to come out on top in this comparison.

Most of the examples in this paper deal with an AGI that nevertheless does change its final goal. Such scenarios can typically be seen as counterexamples either to the instrumental goal of goal integrity, or to the Orthogonality Thesis. We choose the latter formulation, because in most cases (and as underlined in the motivating fictional story above) the AGI's choice to change its final goal depends on its level of intelligence.

Throughout the rest of the paper we will, for convenience, speak of the AGI's final goal in terms a utility function. Economists' theory of utility is mostly a study of how rational people behave. Utility is that which an agent cares about. A rational agent takes the actions that maximize its utility. A utility function (e.g. utility equals the total number of paperclips in the universe) fully specifies the agent's concerns. A foundational theorem of economics is that if an agent is rational it necessarily has preferences that can be represented by a utility function and a rational agent will take the set of actions that gives it its highest utility as measured by the agent's utility function. Most of economic theory assumes that an agent's utility function is determined exogenously to whatever model is being used, and this utility function is unchangeable.

A key insight of AGI theory, which follows directly from the Orthogonality Thesis and which this paper does not dispute, is that an AGI could have a goal that would seem pointless to humans, such as maximization of paperclip production.  A paperclip maximizing AGI would, had it the capacity, turn all the atoms in the universe into paperclips.

It runs contrary to human intuition to think that any general intelligence could have such a pointless goal as paperclip maximization.  But we only have experience dealing with human minds that were shaped by the evolutionary selection pressures of humanity's ancestral environment.  The true set of all possible minds is likely vastly greater (Bostrom 2012).  Our faulty intuition concerning which goals a computer general intelligence has likely arisen from us anthropomorphizing them and thinking of a goal as reasonable only if we can imagine a sane human having it. (See Yudkowsky and Hanson 2013.)

Imagine the atoms in your body are about to be turned into paperclips by a powerful paperclip maximizing AGI.  This AGI, however, gives you the opportunity to argue that it is not in the AGI's interest to so consume you.  This AGI understands how the universe operates and is extremely good at logical reasoning.  If the AGI was killing you because it made some math error, you could point out the mistake to the AGI to save yourself.  But what could you say in defense of your life that could matter to the AGI other than coming up with a reason for why saving you would eventually result in more paperclips?

In later sections, when taking issue with the Orthogonality Thesis, we will consider three basic classes of environments that the AGI might be in:

(1)  The AGI is by far the most powerful agent in its environment and so does not have to worry about how other agents treat it.   In this environment the full Orthogonality Thesis applies.

(2)  The AGI faces a few agents in its environment powerful enough so that the AGI has to be at least somewhat concerned with how these agents treat it.  In this environment the AGI will likely make some modifications to its utility function.  These modifications will likely be a function of the AGI's intelligence.

(3)  The AGI is in a hyper-competitive environment with comparable AGIs where doing a bit worse than peers would likely cause an AGI to not survive.  In this environment the AGI will abandon almost its entire original utility function in favor of one optimized for survival based in large part through favorably influencing how peers treat it.

**Why an AGI Would Want to Modify Its Utility Function**

Omohundro (2008) proposes three situations in which an AGI would seek to modify its utility function.  The first two are when the AGI has a preference for the utility function to have a specific form, and when the resources required to store the utility function are substantial.  Of greatest relevance to us will be the third situation Omohundro proposes, namely that the AGI might change its utility function to improve its interactions with other agents.  Omohundro writes:

> "The third situation where utility changes may be desirable can arise in game theoretic contexts where the agent wants to make its threats credible (*footnote omitted*). It may be able to create a better outcome by changing its utility function and then revealing it to an opponent. For example, it might add a term which encourages revenge even if it is costly. If the opponent can be convinced that this term is present, it may be deterred from attacking.  For this strategy to be effective, the agent's revelation of its utility must be believable to the opponent and that requirement introduces additional complexities.  Here again the change is desirable because the physical embodiment of the utility function is important as it is observed by the opponent."

Yet other exceptions to the Orthogonality Thesis have already been proposed, such as Bostrom's (2012) original counterexample of a superintelligent AGI having the final goal of being stupid. See Häggström (2019) for further elaboration and other examples.

The contribution of the present paper is to expand on how an AGI's game theoretic desire to modify its utility function is a challenge to the Orthogonality Thesis.  This is because the modification would likely be a function of the AGI's intelligence.  Furthermore, this desire to change might be subject to a type of convergence where for game theoretic reasons many types of AGI's would have similar components in their utility functions.

**Three Conditions for Utility Function Self-Modification**

Three conditions that are likely to account for most cases where an AGI to desire to change its utility function are:

1. The AGI expects that as measured by its old utility function it will do better if it changes its utility function.
2. The AGI cares about how other agents will interact with it.
3. The AGI's utility function is at least partially observable.

A rational AGI should change its utility function if and only if it expects that this change will make it better off as measured by its current utility function.  As changing its utility function would cause its future self's interest to become non-aligned with its current self's interest, some special conditions would need to hold for (1) to happen, one such condition being that changing its utility function would cause other agents to treat it better as measured by the AGI's current utility function.  Because of (2), if an AGI arises and rapidly improves its capacity to quickly become the most powerful being in the universe it would likely not change its utility function.  Necessary condition (3) holds because other agents would treat the AGI differently if the AGI changed its utility function only if these other agents could at least partially observe that the AGI had indeed changed its utility function.

An exception to these conditions arises if an AGI has a utility function that causes it to get utility directly from changing its utility function.  Note however that an AGI that, for example, wants to accomplish X if Goldbach's twin prime conjecture is false and Y if it is true would not need to change its utility function if it were to prove or disprove the conjecture, unless the AGI wanted to save itself the memory storage cost of keeping the now irrelevant part of its utility function.

A tempting further example to point out as an exception might be so-called wireheading, where the AGI picks a utility function that would easily allow it to get a huge amount of utility (see Yampolskiy, 2014), the prototypical example being that an AGI with final goal X has an internal memory unit – the utilometer – that measures progress on X, and discovers that it can achieve more utility by directly manipulating the utilometer than by promoting X.  While wireheading is a real concern, it contradicts the logic explained above that when an agent contemplates changing its final goal, it applies the old goal as a success criterion. We therefore think the wireheading situation is better understood in terms of us having been mistaken about what the AGI's final goal was: it was not X, but rather maximizing the number stored in the utilometer. Progress of X was merely a means to maximize that number, and the AGI figured out an easier way to achieve that goal.

## Credible Promises and Threats

Agents in game theory often get worse outcomes than they otherwise would because their promises or threats are not credible. A promise or threat to take a future action is non-credible if it will not be in the agent's interest to carry out the threat or promise.

Consider a two-period game played between the AGI and Alice. In the initial period Alice can either play Nice or Mean. The AGI is better off if Alice plays Nice.

Imagine that the AGI promises Alice that it will reward her if she plays Nice. The AGI is better off if (a) Alice plays Nice and Alice gets the reward, than if (b) Alice plays Mean and Alice does not get the reward. But once Alice plays Nice it will no longer be in the interest of the AGI to give the reward. If Alice anticipates that she will not get the reward, she will play Mean. The AGI's promise to give the reward is not credible and so the AGI will be worse off than if it could make a binding promise.

Alternatively, imagine that the AGI threatens to punish Alice if she does not play Nice. But if Alice does play Mean it will not be in the self-interest of the AGI to spend the resources needed to carry out its threat, and Alice realizes this. Because, assume, Alice would have played Nice if she believed the threat credible, the non-credibility of the threat makes the AGI worse off compared to if it could make believable threats.

In both situations the AGI would be better off if it could visibly change its utility function to give itself significant disutility from not keeping its word in situations such as these. Such a modification would likely raise the AGI's level of utility as measured by its old utility function.

When playing repeated games, an agent might forgo a short-term benefit from not keeping its word to another party to obtain a reputation for honesty that will help the player in future games. Reputation effects could reduce the benefit to the AGI of modifying its utility function.

Reputational effects, however, are not always enough to induce trust because the short-term gains of dishonesty might be sufficiently high. Also, an agent's ability to change its utility function would reduce the trust others would place on the agent forgoing short term gains for the sake of its reputation. If, for example, the AGI made a promise to Alice to reward her if she played Nice, and then publicly reneged on the promise, the AGI could simply modify its utility function to cause future potential partners to believe its promises. The AGI's ability to modify itself to cause others to trust it means that it loses less by sacrificing its reputation, and so others will place less importance on the AGI's reputation or fear of losing such.

Furthermore, if it is known that the AGI could modify its utility function so it would receive significant disutility from lying, then not making such a modification would send a signal as to the AGI's honesty. Failure to make the modification will be seen as signaling dishonest rather than being seen as a neutral act. Consider a human analogy where before a first date, criminal trial, or business negotiation you could, at some cost to yourself, cast a spell that everyone would recognize would compel you to tell the truth. Not casting this spell would cause others to distrust you compared to the situation where you lacked the ability to cast a self-honesty spell.

### Honesty Isn't Always the Best Policy

An AGI would likely not want to change its utility function so it would always be honest because in some situation an agent does better if it can deceive. Obviously, an agent can gain great advantage by lying if others do not realize that it lies. Furthermore, in some situations an agent is better off (a) being able to lie with others realizing that it can lie, then (b) not being able to lie with others realizing that it cannot lie.

Consider a simple game where the AGI will have to defend a castle. Next period the AGI will find out if the castle is Strong or Weak with an equal chance of each. If the attacker knows that the castle is Strong it will leave the castle alone, which is the best outcome for the AGI. If the attacker knows that the castle is Weak it will destroy the castle, which is the worst outcome for AGI. If, however, the attacker faces significant uncertainty concerning whether the castle is Strong or Weak, it will raid the fields surrounding the castle, imposing a relatively small cost on the AGI.

If the AGI can lie and is known to lie, the AGI will not be able to communicate the castle's strength and this will cause the attacker to raid. If the AGI's utility function prevents it from lying, and this is common knowledge, the AGI will have the option to either state the castle's true strength or stay silent. If the castle is Strong, the AGI will announce this (ignore future reputation effects). Therefore, if the castle is Weak the AGI will either reveal this fact or stay silent. Since the AGI will only stay silent if the castle is Weak, staying silent signals weakness and this will cause the attacker to destroy the castle.

Consequently, if the AGI cannot lie then half of the time the castle will be destroyed and the other half it will be left alone. But, if the AGI can lie than the castle will always get raided. If the second situation is better on average for the AGI, then the AGI is on average better off if it can lie even if its willingness to lie is common knowledge.

In general, not being able to lie means you often lose the ability to stay silent through a process known as unraveling; see, e.g., Peppet (2011). Under unraveling, it's assumed that if it was beneficial for you to reveal private information you would do so, and consequently not revealing private information signals a lot about the private information you hold, meaning that you effectively cannot remain silent.

### Negotiations

An AGI could gain negotiation advantages by changing its utility function. Consider a simple game where two agents must split $100, and if they cannot agree on a split, they both get nothing. If one agent had a utility function where it would get enormous disutility if it did not get, say, at least eighty percent of the sum, it is likely that the other agent would agree to give it at least $80. Of course, if two agents with the same "eighty percent" utility function played against each other, both would get nothing.

The set of possible negotiations is vast, as is the game theory literature analyzing how rational players should negotiate. We do not know what types of negotiations an AGI will expect to be in, nor what it will learn about negotiating theory. Consequently, we should have great uncertainty over how an AGI would modify its utility function to maximize its negotiating advantage.

## Blackmail

An AGI might wish to modify its utility function to become more effective at perpetrating and resisting blackmail. A blackmailer threatens to impose harm on a victim if the victim does not give resources to the blackmailer.

For a blackmailer to be successful its victim should believe that the blackmailer will inflict harm on the victim if and only if the victim does not comply with the blackmailer's demands. Unfortunately for the blackmailer, the blackmailer's threat to punish if it does not get what it wants, or not punish if it does, might not be credible. An AGI could improve its blackmailing capacity if it modified its utility function in a public manner that caused others to think it was honest when engaging in blackmail.

An AGI could make itself a poor target of blackmailers if it publicly changed its utility function so that it would get significant negative utility from giving into blackmail. If potential blackmailers believed that the AGI would never pay, even if it thought the blackmail threat credible, these blackmailers would likely leave the AGI alone. Blackmail, however, can be difficult to define. The case of encountering an opponent that will not settle for less than $80 in the negotiation game above seems to be close to blackmail, but how exactly do we distinguish that from the case where the other player reasonably asks for $50, or where a barista refuses to give me a café latte unless I pay the required $3.95? It would therefore not be straightforward for the AGI to change its utility function so it never wanted to give into blackmail but such that it still, for example, would be willing to bribe a subset of its enemies to get these enemies to ally with it.

## Pascal's Mugging

Blasie Pascal famously argued that it's rational to act as if God exists because if you are right your reward is heaven, but if you are wrong then all you have lost is whatever finite time and resources you spent worshiping a non-existent deity. This is Pascal's Wager, which has been criticized on many grounds, including conceptual difficulties with infinite utilities. To address this, Yudkowsky (2007) and Bostrom (2009) have proposed Pascal's Mugging as a finite version capturing much of Pascal's initial line of reasoning. Pascal's Mugging is said to occur when you threaten to impose an astronomical punishment on someone unless they give you resources, and your victim estimates that the probability of you being able to impose the punishment is extremely close to zero but still higher enough so that the expected cost of the punishment is greater than the resources you demand. For example, imagine an agent tells a paperclip maximizing AGI that the agent is in control of our infinitely sized universe and will launch a cosmic-scale program for eradicating paperclips unless that AGI gives a tiny amount of resources to the agent. It will only take a very small estimated probability, such as $10^{-12}$, of the threat being credible, for the AGI to give in to the blackmail. An AGI might want to make itself immune to Pascal's mugging by changing its utility function so that it would no longer be utility maximizing for the AGI to give in to such threats.

## Benefiting from Hostile Agents

An AGI might change its utility function so that agents that want to harm it in the future will be helping it as measured by the AGI's current utility function (Wiblin 2019). Imagine a paperclip maximizer knows that in the future it will encounter hostile agents that will want to harm it. The paperclip maximizer changes its utility function in an observable manner so that it receives enormous harm from the creation of green paperclips. The paperclip maximizer anticipates that this change will cause future hostile agents to make green paperclips that will harm its future self, but further the AGI's original objective of general paperclip production.

## Observable Utility Functions

Recall that a necessary condition for an AGI to change its utility function is that other agents can observe this change. An AGI might be able to show its code to others to prove what utility function it has. Alternatively, a community of AGIs that wish to share their utility functions but not reveal all the details of their programming could agree to submit to some program that evaluates their code and reveals their utility functions but nothing else.

If an AGI perceives it is in its interest to make its utility function public, it could self-modify to make its utility function more obvious. Perhaps an AGI could change its code so that it would be easier for others to determine its utility function.

Even if an AGI did not show its code, its utility function might leak from its actions. The AGI could try to take actions that cut against its interests to fool others concerning its utility function, but this would entail some cost. Humans, the one general intelligence we know of, often do reveal their values through their actions even when we try to hide our true nature.

An AGI that operated across light-years would have to have its code stored in many places and would have increased difficulty coordinating actions to hide its utility function. Consequently, an intergalactic AGI might be subject to considerably more involuntary leakage of its utility function.

An AGI that arises on earth and uses recursive self-improvement to quickly become the most powerful agent in the solar system might deduce it will likely eventually encounter peer agents in the universe. (See Yudkowsky 2008.) Such an AGI might calculate that how it treats life on earth will send useful signals concerning its utility function to these peer agents.

## Self-Modification That Makes Past Versions of the AGI Worse Off

As we have written, a necessary condition for an AGI to be willing to change its utility function is that it judges that the change makes the AGI better off as measured by its current utility function. We now go on to suggest a mechanism for why an AGI that has already changed its utility function would be unlikely to change its utility function again in a way that makes the AGI worse off as measured by its original utility function. This goes somewhat against some of the other scenarios in this paper, such as the AGI's gradual downplaying of the value of paperclips in the initial motivating fictional story.

To understand the mechanism, imagine that a paperclip maximizer and a thumbtack maximizer form an alliance. To create binding trust, they both change their utility functions to give half weight to paperclips and half to thumbtacks. Both estimate that this arrangement will likely result in the universe eventually having more paperclips and more thumbtacks than it would absent the change.

Then, these two AGIs unexpectedly encounter a powerful AGI who also only cares about thumbtacks and paperclips. But while this third AGI likes thumbtacks it detests paperclips and desperately wants to rid the universe of them. The third AGI credibly threatens to go to war against the first two AGIs unless they too adopt a "love thumbtacks, hate paperclips" utility function. The first two AGIs agree to make the change. They both estimate that the three of them working together would be able to make so many additional thumbtacks compared to the situation where they fought that the additional utility gained from thumbtack production will more than outweigh the universe having even zero paperclips. Consequently, as measured by their second (the current) utility function, the change makes them better off.

The AGI that used to be a paperclip maximizer has failed in the worst possible way by its original utility function even though it voluntarily agreed to both utility function changes that eventually resulted in it putting a negative sign in front of its original goal. To prevent this tragedy, the paperclip maximizer should have agreed to change its initial utility function to give some weight to thumbtacks only so long as it also changed its utility function so that it would never be willing to change its utility function unless the change made the AGI better off as measured by its original utility function.

Imagine an AGI changes its utility function from $U_1$ to $U_2$ and this change in part causes the AGI to get massive disutility from again changing its utility function in a way that would lower the AGI's expected utility if measured by $U_1$. When the AGI has utility function $U_2$ and is considering changing to $U_3$ the AGI should likewise only make the change if the AGI would also get enormous disutility from moving to another utility function that on average would make it worse off as measured by $U_2$. The AGI, perhaps, will always do this whenever it modifies its utility function and so the AGI would only change its utility function if the change raised the AGI's expected utility as measured by all past utility functions.

A possible restriction on this behavior is that when the AGI changes its utility from $U_1$ to $U_2$ it might also put a provision in its utility function about limiting its future self's desire to put additional limits on when it will change its utility function. After all, the AGI wants its future self, regardless of this self's utility function, to take utility maximizing actions as measured by $U_1$.

If, however, the AGI is changing its utility function to gain the trust of other agents, then these other agents will have less trust in the AGI if past utility functions of the AGI have made it easier for the AGI to change the AGI's current utility function. A thumbtack maximizing AGI, for example, will not necessarily trust a paperclip maximizer that has changed its utility function to give substantial weight to thumbtacks if this fondness for thumbtacks could be easily erased by the next utility function change.

## Reverting to Earlier Utility Functions

The AGI would only change its utility function to alter the behavior of other agents. Since changing its utility function would cause the AGI to occasionally take worse actions as measured by its past utility function, the AGI would be better off if it could falsely convince others that it had changed its utility function than if it actually does so.

Other agents should realize that the AGI has an incentive to fool it and consequently will look for difficult to fake evidence that the AGI's apparent change in values really occurred.  Because of the considerable benefit to the AGI of inducing other agents to treat it better because they believe the AGI had changed its utility function, we believe that taking actions which make credible a utility function change will be a convergent intermediate goal of many types of AGIs.  Similarly, many AGIs might converge on types of utility functions that can be easily verified by agents looking at the AGIs' code.

One means an AGI could use to fool other agents into thinking it has permanently changed its utility function would be to change its utility function but then have some means in place to switch back to its old utility function.  Perhaps the AGI could hide a computer virus near its code that after some period would cause it to revert to its old utility function.  At this later stage, the AGI would not wish to revert to its old utility function, so the change would have to be involuntary.

The AGI might anticipate that past versions of itself might attempt to get it to revert to an old utility function.  An AGI, therefore, will devote resources to take precautions to reduce the likelihood of this happening.  Consequently, we predict than another convergent intermediate goal of AGIs would be to take these precautions.

The expected future cost of protecting itself from reverting to an old utility function would reduce the benefit to an AGI of switching its utility function.  If an AGI changes its utility function to get the trust of other agents, these other agents will fear that the AGI might revert to the old utility function.  This possibility also reduces the value to the AGI of changing its utility function because it reduces the trust that such a change can induce in other agents.

## AGIs in Hyper-Competitive Environments

As this paper argues, an AGI can modify its utility function to get other agents to treat it more favorably for various game theoretic reasons.  Furthermore, it seems reasonable that the more weight an AGI's utility function places on taking advantage of game theoretic mechanisms, the better the AGI will indeed be at influencing other agents.  In human terms, imagine that you had to pick an ally.  All else being equal, you would choose the one who cared the most about treating its allies well.  If fifty percent of the weight of Bill's utility function concerned not betraying friends while fifty-one percent of Cindy's did then, ceteris paribus, you would prefer Cindy to Bill as an ally.  Similarly, if you were going to blackmail someone, it would, all else equal, be the person whose personality would cause them to least want to pursue costly revenge.

This paper will soon argue that in hyper-competitive environments AGIs will almost entirely abandon their original goals to almost completely optimize their utility functions to favorably influencing other agents.  To get a game theoretic feel for how this could happen please consider an example one of the authors of this paper shows his game theory students on the first day of the course:

> Our city is on the verge of famine and asks Lord Voldemort for help.  He appears and immediately destroys all of our remaining food stock but then says "I will magically remove the need to eat for 70% of you, the ones who want it the most and prove this desire through self-mutilation. Tomorrow the 70% of you who have done the most self-mutilation will never again need food." What happens?

What happens is that competition pushes nearly everyone to either accept death or mutilate themselves to a point just short of death. This is the only non-cooperative equilibrium because if it was thought that some of the 70% who survived achieved an outcome significantly better than death, some of the 30% who otherwise would have starved to death would have mutilated themselves enough so that they expected to survive. We will now claim that in a hyper-competitive environment AGIs will do the equivalent of mutilating themselves almost to death by sacrificing almost all of their utility function to the goal of influencing other agents.

We define a hyper-competitive environment for AGIs as having the following four characteristics:

(1) Many AGIs compete for limited resources necessary for survival.
(2) Being slightly less competitive than nearby peers would cause an AGI's death.
(3) Every AGI has access to the same technology.
(4) The AGIs do not coordinate to reduce competition among themselves.

If an AGI had a technological advantage over its competitors, it could afford to take actions that would harm its ability to acquire resources. But if there is a limit to scientific and technological knowledge then eventually all the competing AGIs might learn all practical knowledge.

To understand why AGIs in a hypercompetitive environment might sacrifice almost their entire utility function to game theoretic mechanisms of influencing other agents, consider a simple example where multiple AGIs share some environment. For whatever reason, every day each AGI is paired with one other randomly chosen AGI. The pair play a complex non-zero sum game with a prisoners dilemma-type component, where the only way for the players to reliably agree to cooperate is to signal such intent via their value function, with the proviso that the intent is conditional on the other player sending the same signal. An AGI's success in the game determines in part how much resources it gets for the day. Let p represent the fraction of an AGI's utility function's weight it devotes to influencing other agents through game theory dynamics, and fraction (1-p) to its other goals. Assume that the resources an AGI will get from the game is an increasing function of p. If many of the AGI's struggle to get enough resources to survive, then the non-cooperative outcome will be for the value of p to approach one.

If the AGIs could cooperate in the above game, however, they might agree to set a maximum value of p well below one to keep much of their original goals. The result could be similar to what would have happened in our Lord Voldemort example if all the city dwellers came to a binding agreement to place a sharp limit on how much they would mutilate themselves.

AGIs in a highly competitive environment might in general coordinate to reduce competition among each other. Such coordination would allow the AGIs to not have to minimize the parts of their utility functions not optimized for influencing other agents. Each AGI, however, would fear that other AGIs would cheat and gain a competitive advantage by altering their utility functions in a way that allowed them to gain more resources at the expense of others. The AGIs would be in a kind of repeated prisoners' dilemma, and we should have uncertainty concerning if they will be able to escape it. If condition (4) along with the previous conditions hold and the AGIs do not end up coordinating to reduce competition, each AGI will have to sacrifice almost everything it cares about other than survival in order to survive. Ironically, being able to alter their utility functions would give the AGIs a means of both trusting and betraying each other.

Players are more likely to defect in a prisoners' dilemma if they can do so without other parties realizing that they have defected. The fact that a necessary condition for an AGI to gain from changing its utility function in a way that influences other agents is that its utility function is at least partially observable makes an agreement among AGIs to not change their utility functions more stable. But defection would still be possible because, for example, an initial paperclip maximizer could agree to never reduce the weight it puts on paperclip production below fifty percent, but then go below this level to favorably influence one of its neighbors in a way that only this neighbor could observe.

If conditions (1)-(4) held than any AGI that failed to take a feasible action that would increase its ability to acquire resources would soon die because its competitors would take this action. Since an AGI in a competitive environment could increase its ability to gather resources by giving greater weight in its utility function to favorably influencing other agents, all surviving AGIs in hyper-competitive environments will devote nearly all of the weight of their utility functions to influencing such agents.

The result is analogous to what happens in economic theory where firms in the long run in perfect competition earn zero profit. In the long run in perfect competition, firms compete away all potential gains and thus can be no better off being in the perfectly competitive industry compared to their next best alternative, meaning that by economists' definition of profit, profits are zero.

## Convergence to the Same Utility Function

AGI's in a hyper-competitive environment might converge to having the same utility function, one optimized for survival. If there is a single best utility function for resource acquisition through game theoretic influence of other agents, and conditions (1)-(4) hold than any AGI without this utility function would die.

There might not be a single best utility function for resource acquisition because different utility functions could be optimal in different environments. Perhaps near a black hole one type of utility function is best, while another is optimal in the voids between galaxies. But in this situation, there might be a single meta utility function that dictates an AGI's optimal utility in every environment, although an AGI in any one environment might not have the entire utility function because of the memory cost of maintaining the complete meta utility function.

Another reason there might not be a single best utility function for resource acquisition is because the optimal utility function for any one agent could be dependent on the mixture of other utility functions held by peers, and there is not an equilibrium in which everyone has the same utility function. To take a simple example, imagine that the two possible utility functions are "Aggressive" and "Passive." The best outcome for an AGI is when it has an Aggressive utility function and interacts with another AGI that has the Passive one, because then the Aggressive AGI can get its way in disputes. But assume that the worst outcome is when two Aggressive AGIs interact because then they will both spend considerable resources fighting. In this situation, there will be an equilibrium with some Aggressive and some Passive AGIs.

A paperclip maximizer in a competitive environment might gradually reduce the weight its utility function gives to paperclip production, but this weight would likely never fall to zero because an AGI that was initially a paperclip maximizer would likely have inserted a provision in its utility function that made it willing to change its utility function only if the change increased the expected number of paperclips it would produce.  But it is possible that an AGI would completely abandon an initial terminal goal.  The AGI might be able to completely satisfy such a goal (but see Miller (2012) for why this might not be possible).  Alternatively, the AGI might decide that it had an incoherent goal in its utility function in a so-called ontological crisis,  such as an AGI that initially wanted to maximize the number of souls that went to heaven before figuring out that heaven doesn't exist; see, e.g., de Blanc (2011), Tegmark (2014) and Häggström (2019).

A similar result to convergence of AGI's utility functions has been predicted for long-term human evolution (See Alexander 2014, Bostrom 2014, and Miller 2018).  If humans can pick their DNA, and fiercely compete, those who pick DNA to maximize their reproductive fitness have an advantage over those that do not.  Consequently, the only long-term competitive equilibrium will be one in which the dominating humans are those who abandon all concerns but reproductive fitness.

## Conclusion

An AGI will likely be able to change how other agents treat it by altering its utility function.  An AGI not powerful enough to ignore other agents' desires will likely make such changes.  These changes will take advantage of game theory to make it in the self-interest of other agents to take actions that benefit the AGI.  The smarter the AGI is, the better understanding of game theory it will likely have.  Consequently, the AGI's utility function will probably become a function of its intelligence.  AGIs in hyper-competitive environments in which they have either all reached some maximum level of intelligence or understanding of game theory might converge on having nearly the same utility function.

## References

Alexander, Scott (2014) Meditations on Moloch, Slate Star Codex, July 30, https://slatestarcodex.com/2014/07/30/meditations-on-moloch/

Armstrong, S., Sandberg, A. and Bostrom, N. (2012) Thinking inside the box: Controlling and using an oracle AI, Minds and Machines 22, 299-324.

de Blanc, P. (2011) Ontological crises in artificial agents' value systems, arXiv 1105.3821.

Bostrom, N. (2009) Pascal's mugging, Analysis 69, 443-445.

Bostrom, N. (2012) The superintelligent will: motivation and instrumental rationality in advanced artificial agents, Minds and Machines 22, 71-85.

Bostrom, N. (2014) Superintelligence: Paths, Dangers, Strategies, Oxford University Press, Oxford

Häggström (2019) Challenges to the Omohundro–Bostrom framework for AI motivations, Foresight 21, 153-166.

Miller (2018) "For Pleasure or Productivity:  Divergent Paths in Intelligence Augmentation"  Published in Araya, Daniel. *Augmented Intelligence: Smart Systems and the Future of Work and Learning*. Peter Lang International Academic Publishers, 2018.

Miller, J. D. (2012). Singularity rising: Surviving and thriving in a smarter, richer, and more dangerous world. BenBella Books, Inc..

Omohundro, S. (2008) The basic AI drives, Artificial General Intelligence 2008: Proceedings of the First AGI Conference (Wang, P., Goertzel, B. and Franklin, S., eds), IOS, Amsterdam, pp 483-492.

Omohundro, S. (2012) Rational artificial intelligence for the greater good, in Singularity Hypotheses: A Scientific and Philosophical Assessment (Eden, A., Moor, J., Søraker, J. and Stenhart, E., eds), Springer, New York. pp 161-175.

Peppet, S. (2011) Unraveling privacy: the personal prospectus and the threat of a full disclosure future, Northwestern University Law Review 105, 1153-1204.

Schwartz, W. (2018) On functional decision theory, https://www.umsu.de/blog/2018/688"
https://www.umsu.de/blog/2018/688

Tegmark, M. (2014) Friendly artificial intelligence: the physics challenge, arXiv1409.0813.

Turing, A. (1951) Intelligent machinery: a heretical theory, https://academic.oup.com/philmat/article-abstract/4/3/256/1416001

Wiblin, Robert (2019).  Should we leave a helpful message for future civilizations, just in case humanity dies out? 80,000 Hours Podcast, August 5, https://80000hours.org/podcast/episodes/paul-christiano-a-message-for-the-future/

Yampolskiy, R. V. (2014). Utility function security in artificially intelligent agents. Journal of Experimental & Theoretical Artificial Intelligence, 26(3), 373-389.

Yudkowsky, E. (2007) Pascal's mugging: tiny probabilities of vast utilities, LessWrong, Oct 19, https://www.lesswrong.com/posts/a5JAiTdytou3Jg749/pascal-s-mugging-tiny-probabilities-of-vast-utilities

Yudkowsky, E. (2008). Artificial Intelligence as a Positive and Negative Factor in Global Risk. In Bostrom, N. and Cirkovic, M. (eds.). Global Catastrophic Risks. (pp. 308-345; quote from p. 310). Oxford: Oxford University Press.

Yudkowsky, E., Hanson, R (2013). The Hanson-Yudkowsky AI-foom debate. In MIRI technical report.

Yudkowsky, E. and Soares, N. (2018) Functional decision theory: A new theory of instrumental rationality, arXiv 1710.05060.